\documentclass[intlimits,twoside,a4paper]{article}

\usepackage[utf8]{inputenc}
\usepackage[eqsecnum]{cmpj3}

\articletype{Regular article}

\title[Critical behavior of cubic systems with extended impurities]%
{Renormalization group study of critical behavior of $O(m)$ model with cubic anisotropy and extended defects: two-loop approximation}
\author[D. Shapoval, M. Dudka]{D. Shapoval\orcid{0000-0003-1648-2035}\refaddr{label1,label2}\thanks{Corresponding author: \email{shapoval@icmp.lviv.ua}.},
        M. Dudka\orcid{0000-0001-6971-8895}\refaddr{label1,label2,label3}}
\addresses{
\addr{label1} Yukhnovskii Institute for Condensed Matter Physics of the National Academy of Sciences of Ukraine, 79011 Lviv, Ukraine 
\addr{label2}  ${\mathbb L}^4$  Collaboration \& Doctoral College for the Statistical Physics of Complex Systems, Leipzig-Lorraine-Lviv-Coventry, Europe
\addr{label3} Lviv Polytechnic National University, 79013 Lviv, Ukraine
}


\begin{document}

\maketitle

\begin{abstract}

We study the impact of quenched extended defects correlated in $\epsilon_d$ dimensions and randomly distributed in the remaining $d - \epsilon_d$ dimensions on the critical properties of a $d$-dimensional $O(m)$-symmetric system with cubic anisotropy. Exploiting the field-theoretic renormalization-group approach within the minimal subtraction scheme, we compute the two-loop renormalization-group functions. Applying resummation techniques to analyze the asymptotic series, we estimate the marginal values of $\epsilon_d$ above which disorder becomes relevant for the physically important cases $m = 2$ and $m = 3$ in $d = 3$. We report numerical values of the critical exponents for the disorder-induced universality classes.

%
\printkeywords
%
\end{abstract}

\section{Introduction}


The concept of universality is one of the cornerstones of the modern theory of critical phenomena. According to this principle, various physical systems, despite their distinct microscopic origins, demonstrate similar critical behavior in the vicinity of their critical points. This behavior is described by scaling laws with universal critical exponents determined by a small set of global parameters, among which the structure of the order parameter --- specifically its symmetry and number of components --- plays an essential role.

The $d$-dimensional $\phi^4$ field model with an $O(m)$-symmetric order parameter serves as a powerful theoretical tool to investigate critical properties across a wide variety of physical systems, particularly magnets \cite{pelissetto2002}. The description of such a field model with one coupling is well established within the renormalization group (RG) approach. There, the asymptotic critical properties are governed by the stable Heisenberg fixed point (FP) of the RG transformation, describing the $O(m)$ universality class (see \cite{pelissetto2002} for a review). However, in real crystalline materials, some degree of order-parameter anisotropy is frequently present. Taking into account even simple cubic perturbations of the order parameter \cite{aharony1973} can change the critical behavior of the $O(m)$ model.
Adding a new term, invariant under cubic group transformations, to the effective Hamiltonian leads to a field model with two coupling constants, whose RG transformations can also have another stable FP depending on the values of $d$ and $m$. It has been revealed that this cubic FP is stable for systems with a sufficiently large number of order-parameter components ($m > m_c(d)$), determining the cubic universality class, while for $m < m_c(d)$, the isotropic Heisenberg FP is stable, leaving the $O(m)$ critical regime relevant. The quantity $m_c(d)$ is called the marginal dimension \cite{pelissetto2002,aharony1973}.

We focus on the case of a three-dimensional system ($d=3$). Estimates obtained from higher-order perturbative RG series refined by resummation methods yield a marginal dimension $m_c(3)$ slightly less than 3 \cite{carmona2000,Folk2000,Kleinert2001,mn_a,mn_b} (for a full list of references, see \cite{Adzhemyan2019}), placing the physically relevant case $m=3$ within the cubic universality class. However, if the initial parameters of the model Hamiltonian lie outside the region of attraction of the cubic fixed point, a fluctuation-induced first-order phase transition substitutes the continuous transition. Recently, this scenario of the $O(3)$ model with cubic anisotropy was used to explain structural phase transitions from cubic to trigonal and from cubic to tetragonal symmetries in certain perovskites \cite{Aharony2022}. For the three-dimensional $O(2)$ model, cubic anisotropy nevertheless may have consequences for critical behavior: being dangerously irrelevant in this case, it is responsible for different dependencies of critical exponents in the high- and low-temperature phases \cite{Nelson1976,Leonard2015,Chelbicki2019,Chelbicki2022}.

Here, we consider the combined effect of two types of anisotropy: cubic anisotropy in the order parameter space and anisotropy in  physical (Cartesian) space generated by structural disorder in the form of extended defects. Such disorder was introduced by Dorogovtsev \cite{dorogovtsev1980phase, dorogovtsev1980exponents, dorogovtsev1980anisotropy} for the $O(m)$ model as quenched random nonmagnetic impurities that are strongly correlated (extended) in $\epsilon_d$ dimensions and randomly distributed over the remaining $d-\epsilon_d$ dimensions. The dilution of extended impurities is considered to be weak. The pointlike uncorrelated defects are described by the case $\epsilon_d=0$, while extended parallel linear and planar defects correspond to $\epsilon_d=1$ and $\epsilon_d=2$, respectively. Extending $\epsilon_d$ to non-negative real numbers can be interpreted as an effective fractal dimension of a complex random defect system \cite{Yamazaki1988,yamazaki1988crossover}. However, the connection between the analytic continuation to non-integer Euclidean dimensions and a physical fractal dimension is not straightforward \cite{Gefen1980,Wu1987,Holovatch1996}.

Treating $\epsilon_d$ on an equal footing with the deviation from the upper critical dimension, $\epsilon = 4 - d$, static and dynamic critical behavior has been studied using RG methods based on a double expansion in the variables $\epsilon$ and $\epsilon_d$ \cite{dorogovtsev1980phase,dorogovtsev1980exponents,dorogovtsev1980anisotropy,Boyanovsky1982,Prudnikov1983,Lawrie1984}. For simplicity, we call this method as the `double $\epsilon$-expansion'. Similar approaches were applied to investigate non-equilibrium critical relaxation \cite{Fedorenko2004} and finite-size scaling corrections \cite{Zhang1992}. More reliable estimates of critical exponents were obtained by applying resummation techniques directly to the asymptotic perturbative RG series at fixed $d$ and $\epsilon_d$ \cite{Blavatska2002,blavatska2003,Fedorenko2004,blavatska2005,Blavatska2006}.  Monte Carlo simulations of models with randomly distributed parallel linear non-magnetic defects support RG estimates for $\epsilon_d = 1$ \cite{Vasilyev2015}. The case $\epsilon_d=1$ appears to have application for  quantum problems with quenched point disorder \cite{Boyanovsky1982}, where defects are considered to be frozen in the time direction, or they can assumed  to be correlated along frictional imaginary-time direction to describe zero-temperature critical behavior \cite{Schakel1997}, that leads to field theories with strongly anisotropic correlated disorder. 
Considering time or imaginary-time dimension to be $\epsilon_d$, double $\epsilon$-expansion RG framework has been used for some quantum field models with extrapolation of results to physically relevant case $\epsilon_d=1$ \cite{Schakel1997,Kirkpatrick1996,Skryabin2000,Narayanan2000,Yerzhakov2021}.

The influence of extended defects on the critical behavior of the $O(m)$ model with cubic anisotropy has been the subject of several studies \cite{yamazaki1985xy,yamazaki1986static,yamazaki1986firstorder,yamazaki1988crossover,Yamazaki1988}, which predicted the possibility of a disorder-induced critical regime for a certain range of model parameters. However, those analyses relied only on double $\epsilon$-expansions. It is known that the convergence of double series in $\epsilon,\epsilon_d$ is poor. Therefore, the critical properties of the $O(m)$ model with cubic anisotropy and extended defects warrant a thorough re-examination using fixed-dimension resummation techniques. Moreover, a discrepancy between the RG functions derived in Ref.~\cite{yamazaki1986static} and those for the isotropic $O(m)$ system with extended defects in Refs.~\cite{Lawrie1984,blavatska2003} further underscores the necessity of this work.

In this study, we calculate the two-loop RG functions for the $\phi^4$ cubic model in the presence of extended defects. Refining these functions through resummation procedures, we analyze which disorder-induced critical regimes are realized for the physically relevant cases of $m = 2$ and $m = 3$ in a three-dimensional system. The remainder of this paper is structured as follows. Section~\ref{Sec2} discusses the criteria for the relevance of structural disorder in the form of extended defects. Section~\ref{Sec3} introduces the field-theoretical model describing the critical properties of anisotropic systems with extended defects along with its renormalization. Section~\ref{Sec4} presents the RG functions obtained in the two-loop approximation and their FPs within the double $\epsilon$-expansion. In Section~\ref{Sec5}, we analyze these functions using the Chisholm-Borel and Pad\'e-Borel resummation procedures. Finally, Section~\ref{Sec6} summarizes and concludes our study.

\section{Relevance of disorder}\label{Sec2}

We treat extended defects as dilution by quenched pointlike defects that are strongly correlated in $\epsilon_d$ dimensions. In this case, random structural imperfections couple to the local energy density,  leading to quenched local random transition temperature disorder. The effects of such disorder on the critical behavior of the $O(m)$ model in the uncorrelated case are predicted by the famous Harris criterion \cite{harris1974}: disorder leads to a new universality class for a continuous phase transition if the specific-heat critical exponent of the undiluted (pure) system is positive:
\begin{equation}\label{harris_orig}
\alpha_{\rm p} = 2 - d \nu_{\rm p} > 0,
\end{equation}
where $\nu_{\rm p}$ is the correlation length critical exponent of the pure system. For instance, only the scalar order parameter case ($m=1$) yields a positive $\alpha_{\rm p}$ at $d=3$ and is consequently affected by weak pointlike disorder at criticality. 

The Harris criterion can be extended to systems with anisotropic perturbations \cite{Bak1976}. When the anisotropy is relevant, $\alpha_{\rm p}$ in the criterion (\ref{harris_orig}) is replaced by the specific-heat critical exponent evaluated at the stable anisotropic FP. In the cubic anisotropy class considered here, $\alpha_{\rm p} \equiv \alpha_{\rm cub} < 0$; therefore, uncorrelated pointlike disorder does not alter the asymptotic critical behavior of the cubic model in $d=3$. Although two-loop RG studies of the cubic model with uncorrelated disorder suggested the possible stability of disorder-induced FPs \cite{Yamazaki1985a,yamazaki1985random,Lawrie1987}, analyses based on resummed six-loop RG series \cite{Calabrese2002,calabrese2003} indicate that the pure FPs (Heisenberg FP  for $m=2$ and cubic FP for $m=3$) are stable, whereas the disorder-induced FP (of the same origin as in the random Ising model) is unstable.~\footnote{Nonetheless, the authors of Refs.~\cite{calabrese2003,Calabrese2002} did not entirely rule out the stability of this FP because of the large error bars in its stability exponents.} 

As argued in Ref.~\cite{Boyanovsky1982}, the Harris criterion is modified in the presence of extended impurities: such randomness is relevant if
\begin{equation}\label{harris_mod}
\epsilon_d > d - \frac{2}{\nu_{\rm p}}. 
\end{equation}
For uncorrelated pointlike disorder ($\epsilon_d = 0$), Eq.~(\ref{harris_mod}) reduces to the standard Harris criterion (\ref{harris_orig}).

\begin{figure}[!htbp]
	\begin{center}
		\includegraphics[width=.5\paperwidth]{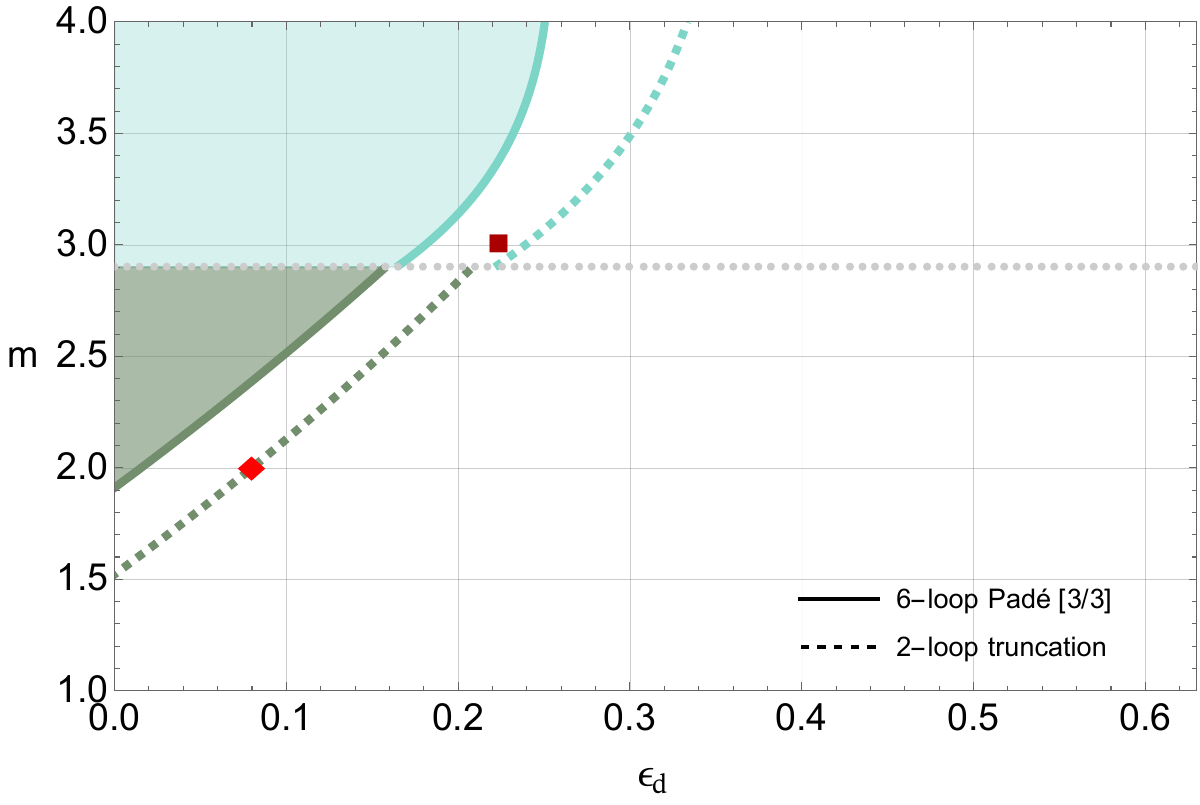}
	\end{center}
\caption{Relevance of quenched disorder in the form of extended defects for the critical properties of the $d=3$-dimensional cubic model, as predicted by modified Harris criterion (\ref{harris_mod}) in the plane $(\epsilon_d,m)$. Horizontal  dotted  grey line marks  estimate of marginal order parameter dimension for three-dimensional cubic model $m_c(3)=2.915$ \cite{Adzhemyan2019}. The shaded regions corresponding to the  critical behavior of the pure Heisenberg model for $m < m(3)$ (dark green) and to the one belonging to cubic universality class  $m>m(3)$  (light green) are separated  from the region where disorder-induced critical behavior is relevant by borderlines $m(\epsilon_d)$, or equivalently $\epsilon_d(m)$, calculated using a Pad\'e [3/3] approximant to the six-loop $\epsilon$-expansions for the correlation-length critical exponents of the pure $O(m)$ model \cite{Kompaniets2017} (solid dark green line) and the cubic model \cite{Adzhemyan2019} (solid light green line). These borderlines therefore indicate the marginal values $\epsilon^{\rm marg}_d(m)$ for disorder relevance. At lower orders of perturbation theory, they shift to higher values of $\epsilon_d$: the dashed curves show the corresponding results based on the two-loop $\epsilon$-expansions. We also present estimates of $\epsilon^{\rm marg}_d(2)$ (red diamond) and $\epsilon^{\rm marg}_d(3)$ (red square) obtained using the fixed-dimension approach for the model (\ref{eff_ham}).}
        \label{marginal_cubic}
\end{figure}

Using the best available high-loop RG results for the critical exponent $\nu_{\rm p}$ of the pure $O(m)$ model, one can examine the relevance of extended defects for $d=3$ across a wider range of $\epsilon_d$. Applying a Pad\'e-approximant [3/3] to the six-loop $\epsilon$-expansion~\cite{Kompaniets2017} and  evaluating it at $\epsilon = 1$, we extract the marginal value $\epsilon^{\rm marg}_d(m)$ above which critical behavior is altered by disorder. This boundary is depicted in Fig.~\ref{marginal_cubic} for all $m < m_c(3)$. For the cubic marginal dimension in $d=3$, we adopt the central $m_c(3)=2.915$ from the  estimate obtained  on six-loop $\epsilon$-expansion basis \cite{Adzhemyan2019}.

Following the arguments of Ref.~\cite{Bak1976}, we can use Eq.~(\ref{harris_mod}) to predict the relevance of extended defects for the cases where the cubic universality class is expected in the undiluted system ($m > m_c(3)$). To this end, we substitute into Eq.~(\ref{harris_mod}) the Pad\'e-approximant [3/3] of the $\epsilon$-expansion for the correlation length critical exponent calculated at the cubic FP \cite{Adzhemyan2019}. Setting $\epsilon = 1$, we extract the corresponding marginal line for this case, which is presented in Fig.~\ref{marginal_cubic}. Since our present study is restricted to the two-loop approximation, Fig.~\ref{marginal_cubic} also displays estimates obtained by directly summing the first three terms of the $\epsilon$-expansions for $\nu_{\rm p}$ and $\nu_{\rm cub}$ in Eq.~(\ref{harris_mod}) within the two-loop approximation.

\section{Field-theoretical model and its RG description}\label{Sec3}

The starting point for our RG study is the effective Hamiltonian of the $d$-dimensional $O(m)$ model with cubic anisotropy and $\epsilon_d$-dimensional defects, extending throughout the system along the coordinate directions $x_{\parallel}$ and randomly distributed in the perpendicular directions $x_{\perp}$. Obtained by averaging over configurations of weak quenched disorder using the replica method~\cite{Emery1975,Edwards1975,Dotsenko01}, this Hamiltonian takes the following form (see e. g.~\cite{yamazaki1988crossover}):

\begin{eqnarray}\label{eff_ham}
{\cal H}_{\rm eff} &=& \sum_{\alpha=1}^{n} \int d^d x\, \Bigg[\frac{1}{2} \left(\mu_0^2{\boldsymbol{\phi}^\alpha}^2 +(\nabla_\perp{\boldsymbol{\phi}^\alpha})^2 + a_0(\nabla_\parallel\boldsymbol{\phi}^\alpha)^2 \right) + \frac{u_0}{4!} \left({\boldsymbol{\phi}^\alpha}^2\right)^2 + \frac{y_0}{4!} \sum_{i=1}^{m}{\phi_{i}^\alpha}^4 \Bigg] \nonumber \\
&+& \frac{v_0}{4!} \sum_{\alpha,\beta=1}^{n} \int d^{\,d}x 
\,d^{\,d}y  \,\, \delta^{d-\epsilon_d}(x_{\perp}-y_{\perp}) {\boldsymbol{\phi}^\alpha}^2(x) {\boldsymbol{\phi}^\beta}^2(y),
\end{eqnarray}
where $\boldsymbol{\phi}^\alpha(x) = \{\phi^\alpha_1(x),\ldots,\phi^\alpha_m(x)\}$ is an $m$-component order-parameter field replicated $n$ times, $r_0$ measures the bare temperature distance  to the critical point $T_c$, $a_0$ is the bare anisotropy constant. The $O(m)$-symmetric bare coupling $u_0$ is positive, the bare coupling $v_0$, being
proportional to minus the variance of the quenched disorder, is negative, and the bare coupling $y_0$ describing the cubic perturbation can have either sign. Note that the quantum model of a disordered antiferromagnet introduced to study the effects of rare locally ordered regions \cite{Narayanan2000} contains terms with the same symmetries as those of the model (\ref{eff_ham}).

The model described by the effective Hamiltonian (\ref{eff_ham}) is known to exhibit rich scaling behavior \cite{yamazaki1985xy,yamazaki1986static,yamazaki1986firstorder,yamazaki1988crossover,Yamazaki1988}. The spatial anisotropy causes the existence of two distinct correlation lengths: one perpendicular ($\xi_\perp$) and one parallel ($\xi_\parallel$) to the direction of the extended defects. As the critical temperature $T_c$ is approached, their divergences are characterized by the corresponding critical exponents  $\nu_\perp$ and $\nu_\parallel$:
\begin{equation}\xi_\perp \sim \vert{}t\vert{}^{-\nu_\perp}, \quad \xi_\parallel \sim \vert{}t\vert{}^{-\nu_\parallel},
\end{equation}
where $t = (T - T_c)/T_c$ denotes the reduced temperature. Consequently, the correlation between order-parameter fluctuations at two different points acquires a spatial anisotropy. Thus, two distinct anomalous dimensions, $\eta_\perp$ and $\eta_\parallel$, characterize the behavior of the correlation function in the directions perpendicular and parallel to the extended defects, respectively.

The critical behavior of the model is analyzed within the field-theoretical RG approach \cite{amit2005,zinnjustin2002,Kleinert2001} using dimensional regularization and the minimal-subtraction scheme ($\overline{\rm MS}$). The renormalized quantities are introduced in the vicinity of the upper critical dimension via renormalizing factors for the field $\phi$, $Z_\phi$, for the couplings, $Z_u$, $Z_v$ and $Z_y$, for the operator $\phi^2$, $\bar{Z}_{\phi^2}$, and for the anisotropy constant, $Z_a$, as follows:
\begin{eqnarray}
    \phi_0=Z_\phi \phi,\, \phi^2_0=\bar{Z}_{\phi^2}\phi^2,\,\, a_0=Z_a a,\, \,u_0=\mu^{\epsilon}Z_\phi^{-2}Z_u u, \,\,v_0=\mu^{\epsilon+\epsilon_d}Z_\phi^{-2}Z_v v,\,\,
    y_0=\mu^{\epsilon}Z_\phi^{-2}Z_y y,
\end{eqnarray}
where renormalization parameter $\mu$ is an external momentum scale.
The corresponding RG functions are obtained from the renormalization constants according to
\begin{eqnarray}
\beta_{g_i}=\mu\frac{\partial g_i}{\partial\mu}\bigg|_{0},
\qquad
\gamma_{\phi}=\mu\frac{\partial\ln Z_{\phi}}{\partial\mu}\bigg|_{0},
\qquad
\bar{\gamma}_{\phi^2} =-\mu\frac{\partial\ln \bar{Z}_{\phi^2}}{\partial\mu}\bigg|_{0},
\qquad
{\zeta}_{a} =\mu\frac{\partial\ln Z_a}{\partial\mu}\bigg|_{0},
\end{eqnarray}
where $g_i=\{u,v,y\}$, while the derivatives are taken at fixed bare parameters.

The asymptotic critical behavior can be governed by a FP $\{g^*\}=\{u^*,v^*,y^*\}$  of RG transformations, defined as common zeroes of the $\beta$-functions:
\begin{eqnarray}\label{fpe}
\beta_{g_i}(\{g^*\}) =0.
\end{eqnarray}
 If there is a stable  FP that is accessible from initial conditions it  corresponds to the critical point of the system. FP stability is  determined from  the corresponding stability matrix
\begin{eqnarray}
B_{ij} = \left. \frac{\partial\beta_{g_i}}{\partial g_j} \right|_{\{g^*\}},
\end{eqnarray}
evaluated at the FP: if  all  its eigenvalues have positive real parts,   the FP is stable.

For the system with extended defects, the critical exponents describing the asymptotic critical behavior along $x_{\perp}$ directions are given by $\gamma$-functions evaluated at the stable FP
\begin{eqnarray} \label{nu_perp}
\nu_\perp^{-1} &= &2-\bar{\gamma}_{\phi^2}(\{g^*\}) -\gamma_{\phi}(\{g^*\}), \\\eta_\perp &=&\gamma_{\phi}(\{g^*\})
\end{eqnarray}
The critical exponents describing the asymptotic critical behavior along the $x_{\parallel}$ directions are calculated on the base of their $x_{\perp}$ counterparts using the $\zeta$-function:

\begin{eqnarray} \label{parallel}
\nu_\parallel = \nu_\perp \left(1-\frac{\zeta_a(\{g^*\})}{2}\right),\qquad \eta_\parallel=2-\left(1-\frac{\zeta_a(\{g^*\})}{2}\right)^{-1}(2-\eta_\perp).
\end{eqnarray}

Other scaling exponents can be calculated from scaling relations. We write them down for susceptibility critical exponent $\gamma$ and the heat-capacity critical exponent $\alpha$:
\begin{eqnarray}\label{gamma}
\gamma &=& (2-\eta_{\perp})\nu_{\perp}=(2-\eta_{\parallel})\nu_\parallel,\\
\alpha &=&2-(d-\epsilon_d)\nu_\perp-\epsilon_d\nu_\parallel.
\end{eqnarray}
We can rewrite Eq. (\ref{gamma}) in the form
\begin{equation}
\label{gamma2}
\gamma^{-1} = 1 - \frac{\bar{\gamma}_{\phi^2}(\{g^*\})}{2-\gamma_{\phi}(\{g^*\})}.
\end{equation}

In the following Section~\ref{Sec4}, we present the RG functions for model (\ref{eff_ham}) and discuss its FPs structure within the double $\epsilon$-expansion.

\section{RG functions and FPs structure in the double $\epsilon$-expansion}\label{Sec4}

Within the above described $\overline{\rm MS}$ renormalization of the model (\ref{eff_ham}), we obtain the RG functions  up to two-loop order in the replica limit $n \to 0$. The explicit expressions for the functions $\beta_{g_i}/g_i$, $\gamma_\phi$ and $\bar\gamma_{\phi^2}$ read:

\begin{eqnarray} \label{bu}
\frac{\beta_u}{u} &=& -\epsilon + \frac{m + 8}{6}u + y + 2v - \frac{3m + 14}{12}u^2 - \frac{11}{6}uy - \frac{5}{12}y^2 \nonumber \\
&& - \frac{uv}{12}\left[\frac{2}{3}(11m + 58) + \frac{m - 4}{3} \frac{\epsilon_d}{\epsilon + \epsilon_d} \right] - \frac{17}{6}vy - \frac{v^2}{144}\left(328 + 32 \frac{\epsilon_d}{\epsilon + \epsilon_d}\right). 
\end{eqnarray}

\begin{eqnarray} \label{bv}
\frac{\beta_v}{v} &=& - (\epsilon + \epsilon_d) + \frac{m + 2}{3}u + y + \frac{4}{3}v - \frac{5(m + 2)}{36}u^2 - \frac{5}{6}uy - \frac{5}{12}y^2 - \frac{7}{6}v^2 \nonumber \\
&& - \frac{1}{18} \left(11 - \frac{\epsilon_d}{\epsilon + \epsilon_d}\right) \left[(m + 2)u + 3y\right]v.
\end{eqnarray}

\begin{eqnarray} \label{by}
\frac{\beta_y}{y} &=& -\epsilon + 2u + \frac{3}{2}y + 2v - \frac{5m + 82}{36}u^2 - \frac{23}{6}uy - \frac{17}{12}y^2 \nonumber \\
&& - \frac{uv}{18}\left[5m + 82 + (m - 4) \frac{\epsilon_d}{\epsilon + \epsilon_d}  \right] - \frac{vy}{12}\left(46 - \frac{\epsilon_d}{\epsilon + \epsilon_d} \right) - \frac{v^2}{18}\left(41 + 4\frac{\epsilon_d}{\epsilon + \epsilon_d} \right). 
\end{eqnarray}

\begin{eqnarray}\label{gap}
\gamma_{\phi} = \frac{1}{72} \left[(m + 2)u^2 + 6uy + 3y^2 + 2(m + 2)uv + 6vy + 2v^2\right].
\end{eqnarray}

\begin{eqnarray}\label{gap2}
\bar{\gamma}_{\phi^2} &=& \frac{m + 2}{6} u + \frac{1}{2} y + \frac{1}{3} v - \frac{m + 2}{12} u^2 - \frac{1}{2} uy - \frac{1}{4} y^2 - \frac{1}{6} v^2 \nonumber \\
&& - \frac{1}{36} \left(6 - \frac{\epsilon_d}{\epsilon + \epsilon_d}\right) v \left[(m + 2)u + 3y\right]. 
\end{eqnarray}

\begin{eqnarray}\label{zeta}
\zeta_a = \frac{v}{3} - \frac{v}{36} \left[(m + 2)u + 3y\right] - \frac{5}{36} v^2. 
\end{eqnarray}
At $\epsilon_d=0$, the functions (\ref{bu})--(\ref{gap2}) agree with the corresponding two-loop expressions for the random-anisotropy model with a cubic distribution of local anisotropy axes in Ref.~\cite{Shapoval2020} (in the case $w=0$ there). In this case, our model describes the cubic model with uncorrelated pointlike disorder \cite{yamazaki1985random,Lawrie1987,calabrese2003}. For $m=2$, the functions (\ref{bu})--(\ref{by}) satisfy the following identities for any $\epsilon_d$:  
\begin{eqnarray}
    \beta_{u}(u + \frac{3}{2} y, v,- y) &=& \beta_{u}(u, v, y) + \frac{3}{2} \beta_{y}(u, v, y), \nonumber \\
   \beta_{v}(u + \frac{3}{2} y, v,- y) &=&  \beta_{v}(u, v, y), \nonumber   \\
   \beta_{y}(u + \frac{3}{2} y, v, - y) &=& - \beta_{y}(u, v, y). \nonumber 
\end{eqnarray}
These identities are also valid in the uncorrelated case~\cite{calabrese2003}. In the absence of cubic anisotropy ($y = 0$), the system reduces to the isotropic $O(m)$ model with extended defects; setting $y = 0$ in the first two $\beta$-functions reproduces the two-loop RG functions of Ref.~\cite{blavatska2003}. 

The FP picture at $\epsilon_d>0$ in the double $\epsilon$-expansion corresponds to that of the cubic model with uncorrelated pointlike disorder in the usual $\epsilon$-expansion \cite{yamazaki1985random,Lawrie1987,calabrese2003}. There are seven conventional FPs, four of which describe the regular undiluted cubic model ($v^* = 0$), these are:
\begin{itemize}
 
\item Gaussian FP $\bf G$ $(u^*=0,v^*=0,y^*=0)$,

\item pure $O(m)$-symmetric Heisenberg FP $\bf PH$ $(u^*\not =0,v^*=0,y^*=0)$,   

 \item Ising FP  $\bf I$ $(u^* =0,v^*=0,y^*\not=0)$,  

\item pure cubic FP $\bf PC$  $(u^*\not =0,v^*=0,y^*\not=0)$.
\end{itemize}

The remaining three FPs are induced by the structural disorder ($v\not = 0$):
\begin{itemize}
\item polymer (self-avoiding walks) FP   $\bf PO$ $(u^* =0,v^*\not=0,y^*=0)$,

\item random  Heisenberg FP $\bf RH$ $(u^*\not =0,v^*\not=0,y^*=0)$,   

\item random  cubic FP $\bf RC$ $(u^*\not =0,v^*\not=0,y^*\not=0)$.  \end{itemize}

Note that for $u = 0$, the last two $\beta$-functions (\ref{bv}) and (\ref{by}) are degenerate at the one-loop level. For such cases, an $O(\sqrt\epsilon)$-expansion was developed \cite{sqrtepsilon_a,sqrtepsilon_b,sqrtepsilon_c}. In our case, it leads to the emergence of  a pair of new FPs:
\begin{eqnarray}\label{ri}
u^{*}=0\qquad
    v^{*} = \mp 3 \sqrt{\frac{6(\epsilon + 3 \epsilon_d)}{53}} \qquad
    y^{*} = \pm 4 \sqrt{\frac{6(\epsilon + 3 \epsilon_d)}{53}}.
\end{eqnarray}
Here, we consider only the physical case of $v<0$; therefore, we restrict our attention to the branch with $v^* < 0$. This FP has the same origin as the corresponding FP of the random Ising model (see e. g. \cite{Folk2003}); thus, we denote it by $\bf{RI}$. 

A similar degeneracy occurs for  all $\beta$-functions at $m=2$; therefore, the FP $\bf RC$ does not appear. Instead, it is replaced by another pair of FPs with three nonzero coordinates:
\begin{eqnarray}
 u^{*} = \pm 6 \sqrt{\frac{6(\epsilon + 3 \epsilon_d)}{53}} \qquad
    v^{*} = \mp 3 \sqrt{\frac{6(\epsilon + 3 \epsilon_d)}{53}} \qquad
    y^{*} = \mp 4 \sqrt{\frac{6(\epsilon + 3 \epsilon_d)}{53}}\label{ri2}
\end{eqnarray}
As in the previous case, we consider only the FP with $v^* < 0$. Because its origin is similar to that described above, we denote this FP by ${\bf RI2}$.  

As can be seen from the picture above, the double $\epsilon$ and $\sqrt\epsilon$-expansions generally give different numbers of FPs in the cases $m=2$ and $m=3$.  For $m=3$ there are 9 FPs: 7 conventional FPs and the pair (\ref{ri}), while for $m=2$ there are 10 FPs: in addition to (\ref{ri}), a new pair (\ref{ri2}) replaces $\bf RC$.

The stability of all these FPs was discussed within the double $\epsilon$- and $\sqrt\epsilon$-expansions in Refs.\cite{yamazaki1985xy,yamazaki1986static,Yamazaki1988} without conclusive results for the physically interesting cases $m=2,\,m=3$. Therefore, in the next Section, we perform analysis of the resummed RG functions at fixed values of $\epsilon_d$ and $d=3$.

\section{Results at fixed $d$, $\epsilon_d$- approach}\label{Sec5}

Since perturbative expansions for renormalization-group (RG) functions are known to be asymptotic series, at best \cite{pelissetto2002,Holovatch02},  we employ resummation techniques to extract reliable numerical estimates from them. 
Because of the shortness of the RG series (\ref{bu})--(\ref{gap2}), our choice of resummation procedures is restricted. Specifically, we utilize the Chisholm-Borel and Pad\'e-Borel methods (for detailed descriptions, see, e.g., Ref.~\cite{Holovatch02}). Here, we present data obtained primarily within the first resummation scheme, drawing upon results from the Pad\'e-Borel procedure where it is needed in our discussion.

Following Ref.~\cite{blavatska2003}, we approximate the Borel images of the functions given in Eqs.~(\ref{bu})--(\ref{by}) using a symmetric three-variable Chisholm approximant with a linear denominator:
\begin{eqnarray} \label{cf} 
\mathcal C_f(u,v,y)= \frac{b_{000}+b_{100}u+b_{010}v+b_{001}y+b_{110}uv+b_{101}uy+b_{011}vy}{1+c_{100}u+c_{010}v+c_{001}y},
\end{eqnarray}
whose coefficients are fixed by requiring that the expansion of $\mathcal C_f$ reproduces the Borel-transformed series (while in the Pad\'e-Borel procedure we approximate the same expressions by a Pad\'e approximant [1/1]). This construction is designed such that if any of the variables ($u, v, y$) vanishes, the approximant correctly reduces to the familiar results for the corresponding reduced models. The resummed functions are subsequently obtained via the inverse Borel transformation. For fixed values of $m$, $\epsilon$, and $\epsilon_d$, we can extract the FPs from Eq.~(\ref{fpe}).  

We are mainly interested in the case $d=3$ ($\epsilon=1$). Prior to presenting our results for $\epsilon_d > 0$, we discuss the outcome for $\epsilon_d = 0$. 
As discussed above, the double $\epsilon$- and $\sqrt\epsilon$-expansions give different numbers of FPs for $m=2$ and $m=3$. Additionally, new solutions for FPs can appear when solving the system of non-algebraic highly nonlinear equations. Moreover, the resummation procedure cannot eliminate all spurious FPs, which are not perturbative in $\epsilon$ and do not coincide with the $\bf G$ at $d=4$ (see discussion e. g. in~\cite{Delamotte2008}). Here, we obtained 10 FPs for $m = 2$ and 9 FPs for $m = 3$. We identify one FP with $(u^*<0,v=0,y^*>0)$ for $m=2$ as unphysical. Then 7 FPs can be associated with the corresponding FPs of the double $\epsilon$-expansion: $\bf G$, $\bf I$, $\bf PH$,  $\bf PC$, $\bf PO$, $\bf RH$, and $\bf RI$. For each considered value of $m$, the two remaining FPs  with all nonzero coordinates are potential candidates for the $\mathbf{RC}$ FP for $m = 3$ or the $\mathbf{RI2}$ FP for $m = 2$. We examined their behavior as $\epsilon \to 0$. In both pairs, the FP with the highest value of $u^*$ flows toward the Gaussian FP $\mathbf{G}$ as $\epsilon$ decreases, whereas the FP with the lowest value of $u^*$ disappears at some small value of $\epsilon$. Because only the Gaussian FP $\mathbf{G}$ determines the long-range properties of the $\phi^4$ model at $\epsilon = 0$, we consider the FPs with the highest $u^*$ within each pair to be the physical FPs (we denote them by $\mathbf{RC}$ for $m = 3$ and $\mathbf{RI2}$ for $m = 2$), while we treat those with the lowest $u^*$ as spurious FPs (artifacts of the resummation procedure), denoting them by $\mathbf{RC^\prime}$ and $\mathbf{RI2^\prime}$, respectively.

Among these FPs, $\mathbf{PH}$ is stable for $m = 2$, and $\mathbf{PC}$ is stable for $m = 3$. Additionally, the $\mathbf{RI}$  appears to be stable in both cases; however, taking into account the results of Refs.~\cite{Calabrese2002,calabrese2003}, we consider the stability of $\mathbf{RI}$ to be an artifact of our low-loop approximation and expect its stability  to change in higher-order loop calculations. The remaining FPs are unstable. 

Below, we analyze the properties of all FPs for $\epsilon_d > 0$ separately for $m = 2$ and $m = 3$.

\begin{figure}[!htbp]
	\begin{center}
		\includegraphics[width=.3\paperwidth]{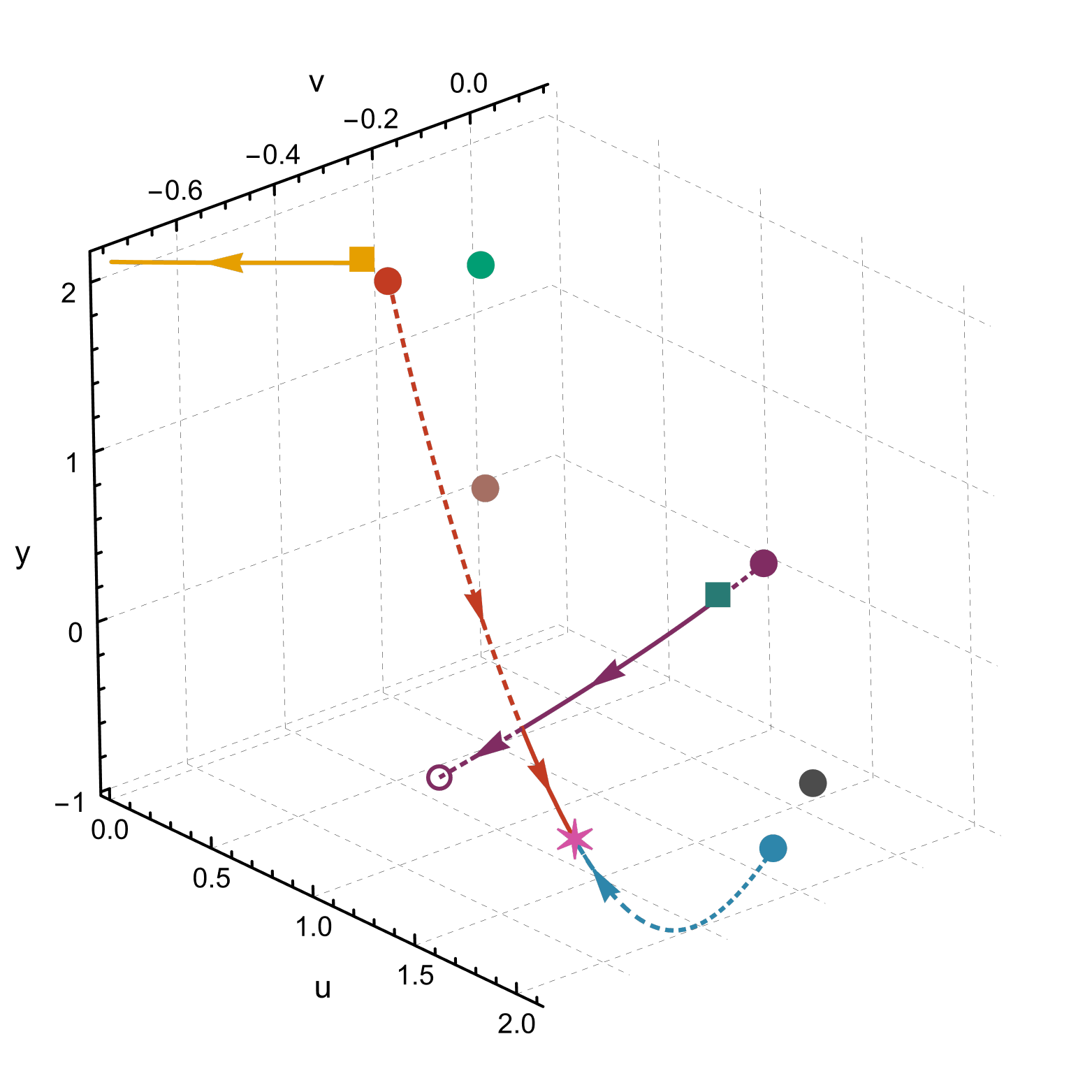}
        \raisebox{0.85cm}{\includegraphics[width=.3\paperwidth]{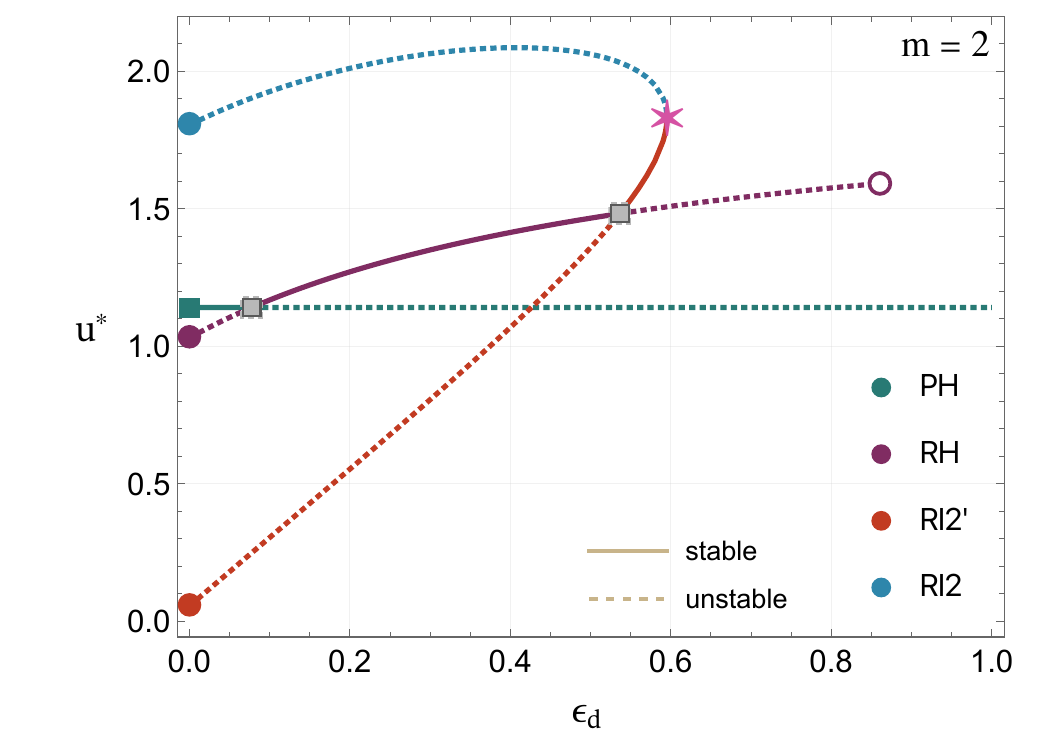}}
	\end{center}
    \caption{FFPs of the RG transformations of the model (\ref{eff_ham}) obtained in the two-loop approximation using Chisholm--Borel resummation for $m=2$ and $d=3$ at different values of $\epsilon_d$.  
Left panel: Tracks of FPs in the coupling space $(u,v,y)$ as $\epsilon_d$ varies from $\epsilon_d=0$ to $\epsilon_d=1$ (arrows indicate increasing $\epsilon_d$). Right panel: The $u^*$ coordinates of FPs involved in bifurcations as functions of $\epsilon_d$. The colors distinguish the $\mathbf{PH}$ (teal), $\mathbf{RH}$ (purple), $\mathbf{RI2}$ (blue),  $\mathbf{RI2^\prime}$ (red), $\mathbf{G}$ (brown), $\mathbf{I}$ (green), $\mathbf{PC}$ (dark gray), and $\mathbf{RI}$ (orange)  branches. Starting locations at $\epsilon_d=0$ for stable (unstable) FPs are shown by squares (circles) of the corresponding colors.
The part of the path traced by each FP while it is stable (unstable) is shown by a solid (dashed) line.
The star marks the saddle-node bifurcation, where $\mathbf{RI2}$ and $\mathbf{RI2^\prime}$ merge and mutually annihilate, while gray squares in the right panel mark transcritical bifurcations where the involved FPs exchange their stabilities. The open circle marks entry into the region of coupling values where poles appear in the resummation.}
        \label{fps3d}
\end{figure}

\subsection{The case $m=2$}

Let us first consider the results for $m = 2$. Our data for the FPs  obtained using Chisholm-Borel resummation for $0\le\epsilon\le 1$ are collected in Fig.~\ref{fps3d}. While the left panel presents locations of the FPs as $\epsilon_d$ is varying, the right panel reports $u$-coordinates of the FPs involved in bifuractions (see below). At small $\epsilon_d$, the $\mathbf{PH}$  is stable, governing the pure $O(2)$ universality class; consequently, weak dilution of extended defects is irrelevant, with no distinction between the parallel and perpendicular directions. As $\epsilon_d$ increases, $\mathbf{RH}$ moves toward  $\mathbf{PH}$ and collides with it  at $\epsilon_d\approx 0.078$, where the FPs exchange their stability.  This is known as  a transcritical bifurcation in bifurcation theory, a mathematical field that studies qualitative changes in the behavior of dynamical systems through changes in the topology of the FPs of the differential equations describing them \cite{Hirsch2013}.
Beyond this point, the stable $\mathbf{RH}$  governs the critical regime of model (\ref{eff_ham}). This picture agrees well with our two-loop estimate of $\epsilon_d \approx 0.08$ obtained from the modified Harris criterion (see Section~\ref{Sec2} and Fig.~\ref{marginal_cubic}). However, we could not track  $\bf RH$ all the way to the physically relevant case of linear defects ($\epsilon_d = 1$). It remains stable with further increase in $\epsilon_d$ until a second transcritical bifurcation occurs involving the $\mathbf{RI2^\prime}$  (at $\epsilon_d \approx 0.537$), where the two FPs exchange stability. Note that the now-stable $\mathbf{RI2^\prime}$ has a complex-conjugate pair of stability exponents, causing RG flows to approach it 
spirally. For larger values of $\epsilon_d$, the stable $\mathbf{RI2^\prime}$ eventually mutually annihilates with unstable $\mathbf{RI2}$ (at $\epsilon_d \approx 0.595$) in a saddle-node bifurcation (see e. g. \cite{Hirsch2013}). Note that results obtained via Pad\'e-Borel resummation predict neither the second transcritical bifurcation nor the saddle-node bifurcation; in that scheme, the $\mathbf{RH}$ remains stable up to $\epsilon_d \approx 0.473$. While it technically remains stable at larger $\epsilon_d$ values within Pad\'e-Borel, it can only be followed continuously using the principal-value prescription.

Therefore, treating the spurious $\mathbf{RI2^\prime}$ as a resummation artifact, we expect the critical behavior of the model (\ref{eff_ham}) at larger $\epsilon_d$ to be described by the $\mathbf{RH}$, although it cannot be reliably accessed within the present loop order. In an opposite case (if we do not take into account FP $\bf {RI}$) no stable FPs remain in the coupling subspace with $u > 0$ and $v < 0$. This is interpreted as a fluctuation-induced first-order phase transition and analyzed in Ref.~\cite{yamazaki1986firstorder} for the model (\ref{eff_ham}).

\begin{figure}[!htbp]
	\begin{center}
		\includegraphics[width=.3\paperwidth]{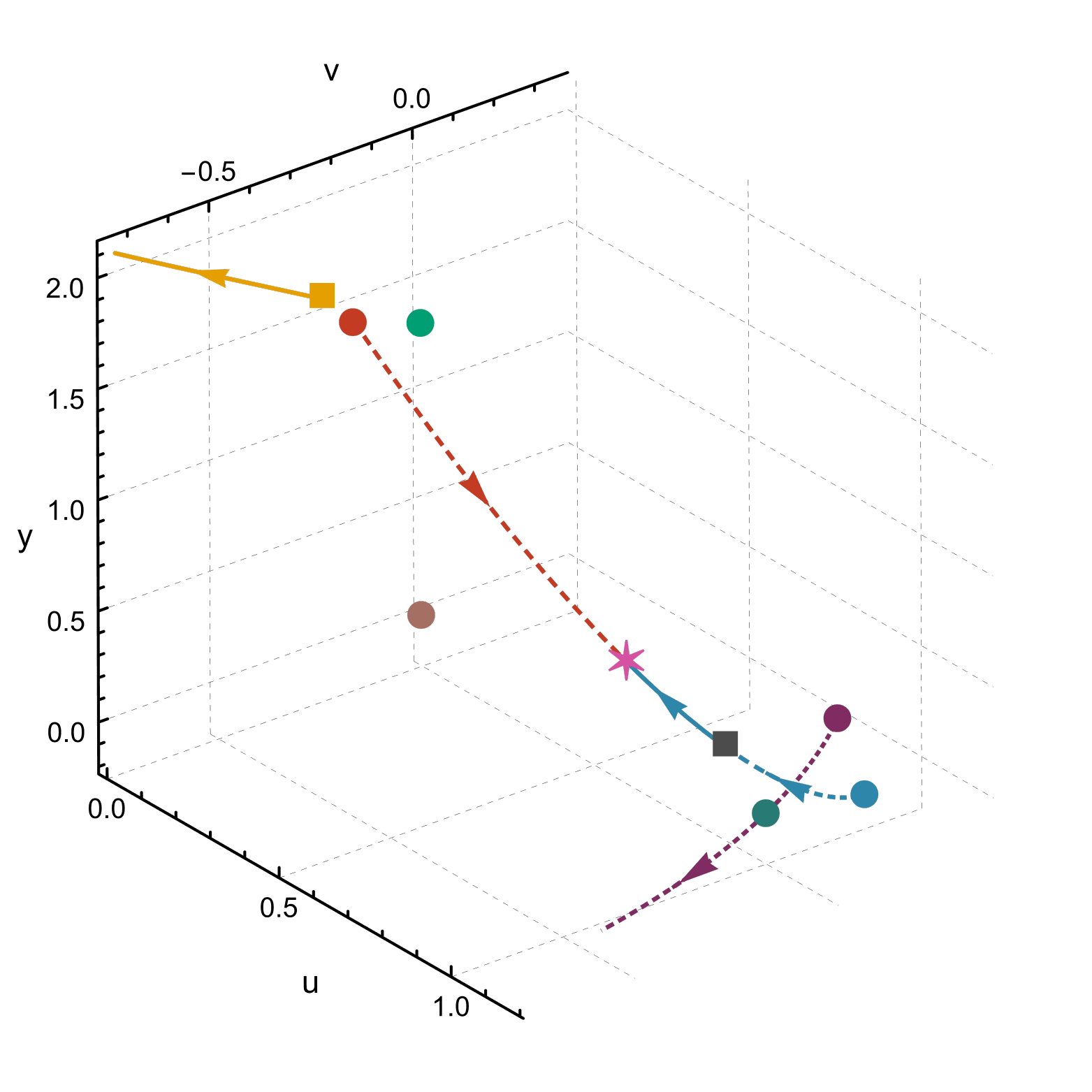}
        \raisebox{0.85cm}{\includegraphics[width=.3\paperwidth]{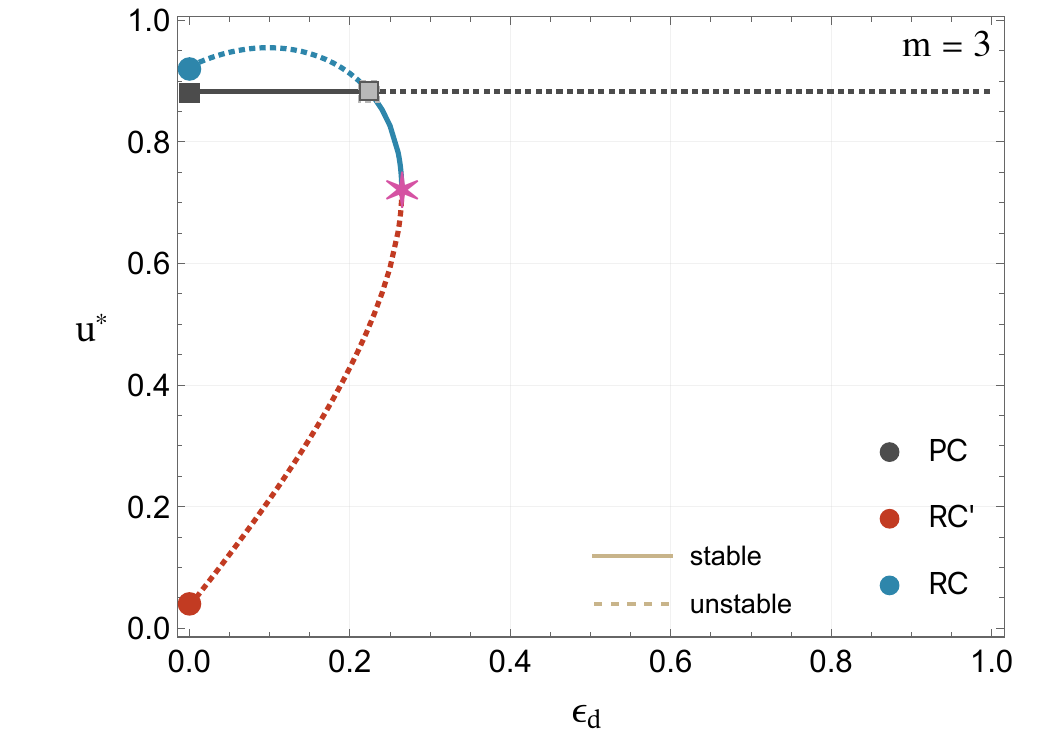}}
	\end{center}
    \caption{FPs of the RG transformations of the model (\ref{eff_ham}) obtained in the two-loop approximation using Chisholm--Borel resummation for $m=3$ and $d=3$ at different values of $\epsilon_d$. Left panel: Tracks of FPs in the coupling space $(u,v,y)$ as $\epsilon_d$ varyies from $\epsilon_d=0$ to $\epsilon_d=1$ (arrows indicate increasing $\epsilon_d$). Right panel:  The $u^*$ coordinates of FPs involved in bifurcations as functions of $\epsilon_d$. The colors distinguish the $\mathbf{PH}$ (teal), $\mathbf{RH}$ (purple), $\mathbf{RC}$ (blue),  $\mathbf{RC^\prime}$ (red), $\mathbf{G}$ (brown), $\mathbf{I}$ (green), $\mathbf{PC}$ (dark gray), and $\mathbf{RI}$ (orange)  branches. Starting locations at $\epsilon_d=0$ for stable (unstable) FPs are shown by squares (circles) of the corresponding colors. The part of the path traced by each FP while it is stable (unstable) is shown by a solid (dashed) line. The star marks the saddle-node bifurcation, where $\mathbf{RC}$ and $\mathbf{RC^\prime}$ merge and mutually annihilate, while the gray square in the right panel marks a transcritical bifurcation where the involved FPs exchange their stabilities. }
        \label{fpsm2m3}
\end{figure}

\subsection{The case $m=3$}
 Our data  for the FPs obtained using Chisholm-Borel resummation for $0\le\epsilon\le 1$ for $m=3$ are collected in Fig.\ref{fpsm2m3}. The left and right panels present the results in a similar way to those for $m=2$.
In this case, for small $\epsilon_d$, the $\mathbf{PC}$  is stable. As $\epsilon_d$ increases, the $\mathbf{RC}$  moves toward $\mathbf{PC}$. Then, at $\epsilon_d \approx 0.224$, a transcritical bifurcation occurs where $\mathbf{RC}$ and $\mathbf{PC}$ collide and exchange their stability. Consequently, for $\epsilon_d > 0.224$, the now-stable $\mathbf{RC}$ fixed point dictates the critical properties of the model (\ref{eff_ham}). This behavior agrees with the results of the modified Harris criterion (see Section~\ref{Sec2} and Fig.~\ref{marginal_cubic}), being  close to the two-loop estimate for the marginal value of $\epsilon_d\approx 0.239$ at which extended defects become relevant to critical behavior. As $\epsilon_d$ increases further, the $\mathbf{RC}$ and $\mathbf{RC^\prime}$ approach each other and eventually mutually annihilate in a saddle-node bifurcation at $\epsilon_d \approx 0.265$. In contrast to the $m = 2$ case, a similar scenario occurs when Pad\'e-Borel resummation is applied, although the saddle-node bifurcation is shifted to a higher value, $\epsilon_d \approx 0.292$.

Therefore, if it is possible to eliminate the influence of the spurious FP $\bf RC^\prime$ on our results, the random cubic FP $\bf RC$ would describe the critical properties of the model (\ref{eff_ham}) for larger $\epsilon_d$. We expect that studying the model at higher loop orders and, consequently, applying more sophisticated resummation techniques will help to rule out spurious FPs.

We also want to note that, for $m=3$, all stable FPs  as stable nodes. Therefore, no spiral approach of RG flows to these FPs is predicted. Moreover, they have a positive coordinate associated with cubic anisotropy: $y^*>0$. This means that they are not accessible from the initial conditions with $y<0$. Therefore, the model (\ref{eff_ham}) with such initial values of the anisotropy coupling should undergo a fluctuation-induced first-order phase transition, similarly to the case of the pure cubic model.

\subsection{The  critical exponents}

\begin{figure}[!htbp]
	\begin{center}
		\includegraphics[width=.3\paperwidth]{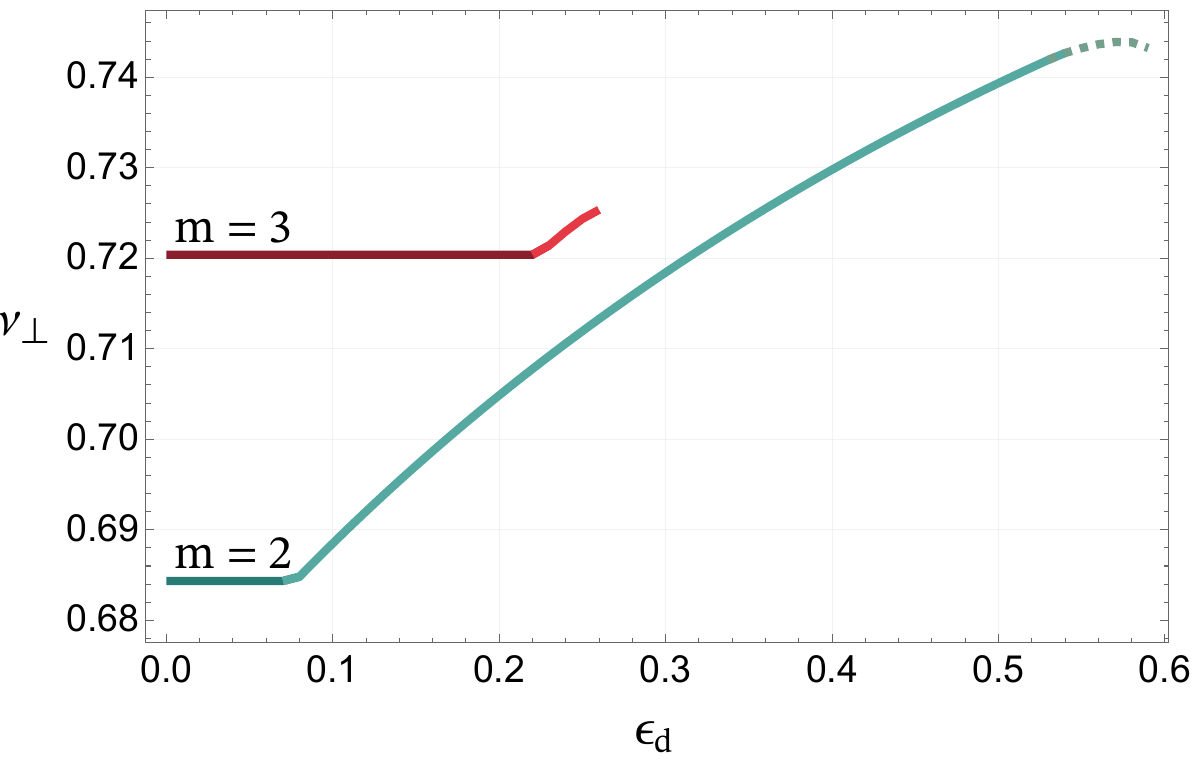}
        \includegraphics[width=.3\paperwidth]{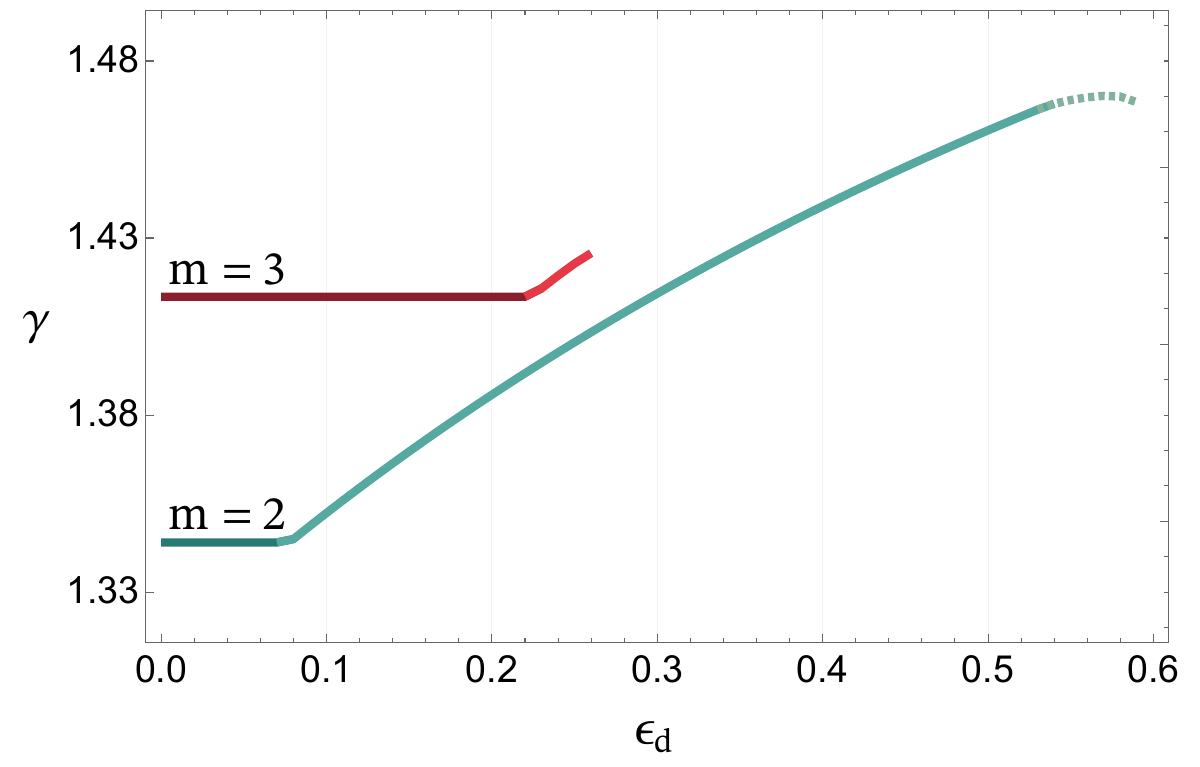}
	\end{center}
	\caption{Critical exponents $\nu_\perp$ (left panel) and $\gamma$ (right panel) as functions of the extended defect dimension $\epsilon_d$ for     $m=2$ and $m=3$ at $d=3$. The two-loop series for $\nu_\perp$ and $\gamma$ are resummed using the Chisholm--Borel method and evaluated at the corresponding stable FPs. The dashed parts are values calculated at a spurious FP for $m=2$.}
        \label{nu_gamma}
\end{figure}

In this subsection we present our results for critical exponents of the cubic model with extended defects at $d=3$. In Fig.~\ref{nu_gamma}  we plot the values of the critical exponents
$\nu_\perp$ and $\gamma$ for the $O(m)$ model with cubic anisotropy for $m = 2$ and $m=3$ as functions of the parameter $\epsilon_d$. They are calculated by substituting the coordinates of the stable FPs into the resummed expressions for the inverse critical exponents (\ref{nu_perp}) and (\ref{gamma2}). Other exponent can be found using familiar scaling relations (\ref{parallel}). For the resummation, we use the Chisholm-Borel procedure with the same approximant (\ref{cf}) used to extract the FPs. As are seen from the figures: the critical exponents remain constant and equal to the corresponding exponents of the pure Heisenberg model (for $m=2$) and the pure cubic model (for $m=3$) until $\epsilon_d$ reaches its marginal value, and only for $\epsilon_d > \epsilon^{\rm marg}_d$ do the values of the exponents increase. Therefore, the presence of extended defects leads to a divergence of the susceptibility that is more sharp than in the pure systems.

We have checked that the results for $\nu_{\parallel}$, not shown here, demonstrate the same tendencies under change of the parameter $\epsilon_d$. Moreover, the relation $\nu_{\parallel} > \nu_{\perp}$ holds for all $\epsilon_d$. This behavior was already observed in studies of the impact of extended defects on the critical behavior of the pure $O(m)$ symmetric model and was explained as follows: extended defects interrupt interactions in the directions perpendicular to them, so fluctuations are stronger in the parallel direction and the correlation length diverges more sharply.

\section{Conclusions}\label{Sec6}

In this study, we investigated how the critical behavior of a $d=3$-dimensional $O(m)$-symmetric model changes  under the mutual influence of two types of anisotropy: cubic anisotropy in the order parameter space and spatial anisotropy in Cartesian space introduced by structural disorder in the form of quenched extended defects. These defects can be considered $\epsilon_d$-dimensional objects that extend throughout the system and are randomly distributed across the remaining $d - \epsilon_d$ dimensions. Previous results obtained for such systems were based on double $\epsilon$-expansions and were thus rather qualitative in character.

The relevance of extended defects is governed by the modified Harris criterion. Combining this criterion with high-order loop $\epsilon$-expansions for the critical exponents of the pure $O(m)$ and cubic models, we predicted the marginal defect dimensionalities $\epsilon_d$ above which disorder becomes relevant in three dimensions for two distinct regimes: $m > m_c(3)$ and $m < m_c(3)$, where $m_c(3)$ denotes the marginal order-parameter dimension of the pure cubic model at $d=3$ and in the highest accuracy $m_{c}(3) \simeq 2.915$ \cite{Adzhemyan2019}.

To verify these estimates, we analyzed an anisotropic $\phi^4$ field theory with three coupling constants using the field-theoretic RG approach. We derived the RG functions for the model with arbitrary $m$ up to two-loop order within the minimal subtraction scheme. Applying Chisholm-Borel (as well as Pad\'e-Borel) resummation techniques, we analyzed the FPs at fixed $\epsilon = 1$ ($d=3$) and varying $\epsilon_d$ for the physically relevant number of order-parameter components, $m = 2$ and $m = 3$. We obtained the marginal values of $\epsilon_d$ at which pure Heisenberg criticality crosses over to the regime governed by the stable random Heisenberg FP ($\mathbf{RH}$) for $m=2$, and at which pure cubic criticality crosses over to the regime governed by the stable random cubic FP ($\mathbf{RC}$) for $m=3$.
We expect these threshold estimates to shift toward lower values in higher-loop calculations.

In contrast to the isotropic $O(m)$ model with extended defects \cite{Blavatska2002,blavatska2003}, we could not track these stable FPs all the way to the physically relevant case of parallel linear defects ($\epsilon_d = 1$) due to the existence of spurious FPs, which we interpret as artifacts of the applied resummation procedures. At certain values of $\epsilon_d$ above the marginal values, these spurious FPs distort the topology of the RG flow, removing the stability of $\mathbf{RH}$ for $m=2$ or annihilating $\mathbf{RC}$ for $m=3$. We have observed  RG flows approach in spirally manner predicted for model (\ref{eff_ham}) \cite{yamazaki1985xy, yamazaki1986static,Yamazaki1988} only for spurious FP for $m=2$. We anticipate that accounting for higher orders of perturbation theory will enable the use of more elaborate resummation schemes capable of eliminating the spurious FPs.

Finally, we calculated the critical exponents for order-parameter dimensions $m=2$ and $m=3$. Both cases demonstrate that when quenched extended defects are relevant, they lead to larger values of the susceptibility critical exponent, consistent with previous findings for the isotropic $O(m)$ model with this type of disorder. We hope that our results will motivate further studies of the influence of extended structural defects on the critical behavior of three-dimensional magnetic systems. In our opinion, Monte Carlo simulations of systems with parallel linear defects seem feasible, as demonstrated by those previously conducted for the Ising model~\cite{Vasilyev2015}.


\section*{Acknowledgements} 
We thank Prof. Yurij Holovatch for reading the first draft of this manuscript, for valuable comments, and for enlightening discussions.
The work was supported by the National Research Foundation of Ukraine under Project 2023.03/0099 ``Criticality of complex systems: fundamental aspects and applications''.
 

\bibliographystyle{cmpj}
\bibliography{references}


\ukrainianpart

\title{Ренормгрупове дослідження критичної поведінки $O(m)$-моделі з кубічною анізотропією та протяжними дефектами: двопетлеве наближення}
\author{Д. Шаповал\orcid{0000-0003-1648-2035}\refaddr{label1,label2}, М. Дудка\orcid{0000-0001-6971-8895}\refaddr{label1,label2,label3}}
\addresses{
\addr{label1} Інститут фізики конденсованих систем імені І.Р. Юхновського Національної академії наук України, 79011 Львів, Україна 
\addr{label2} Спiвпраця ${\mathbb L}^4$ \& Докторський коледж iз статистичної фiзики складних систем, Ляйпцiґ-Лотарингiя-Львiв-Ковентрi, Європа
\addr{label3} Нацiональний унiверситет ``Львiвська полiтехнiка'', 79013 Львів, Україна
}

\makeukrtitle

\begin{abstract}

Досліджено вплив заморожених протяжних дефектів, скорельованих у $\epsilon_d$ вимірах і випадково розподілених у решті $d-\epsilon_d$ вимірах, на критичні властивості $d$-вимірної $O(m)$-симетричної системи з кубічною анізотропією. Використовуючи теоретико-польовий ренормгруповий підхід у схемі мінімального віднімання, обчислено двопетлеві ренормгрупові функції. Застосовуючи методи пересумовування для аналізу асимптотичних рядів, оцінено граничні значення $\epsilon_d$, вище яких безлад стає релевантним для фізично важливих випадків $m=2$ та $m=3$ при $d=3$. Наведено числові значення критичних показників відповідних класів універсальності, зумовлених наявністю безладу.

\tolerance=3000%

\keywords{критичні явища, структурний безлад, ренормалізаційна група, кубічна анізотропія, біфуркації}

\end{abstract}

\end{document}